\documentclass[conference]{IEEEtran}
\IEEEoverridecommandlockouts

\usepackage{cite}
\usepackage{amsmath,amssymb,amsfonts}
\usepackage{algorithmic}
\usepackage{graphicx}
\usepackage{booktabs}
\usepackage{textcomp}
\usepackage[dvipsnames]{xcolor}
\def\BibTeX{{\rm B\kern-.05em{\sc i\kern-.025em b}\kern-.08em
    T\kern-.1667em\lower.7ex\hbox{E}\kern-.125emX}}
\begin{document}

\title{Knowledge Enhanced Multi-Domain Recommendations in an AI Assistant Application}

\author{
    \IEEEauthorblockN{Elan Markowitz\textsuperscript{1,2}, Ziyan Jiang\textsuperscript{1}, Fan Yang\textsuperscript{1}, Xing Fan\textsuperscript{1},}
    \IEEEauthorblockN{Zheng Chen\textsuperscript{1}, Greg Ver Steeg\textsuperscript{1}, Aram Galstyan\textsuperscript{1}}
    \IEEEauthorblockA{\textsuperscript{1}Amazon Alexa AI\\
    \textsuperscript{2}University of Southern California}
}

\maketitle

\begin{abstract}
This work explores unifying knowledge enhanced recommendation with multi-domain recommendation systems in a conversational AI assistant application. Multi-domain recommendation leverages users' interactions in previous domains to improve recommendations in a new one. Knowledge graph enhancement seeks to use external knowledge graphs to improve recommendations within a single domain. Both research threads incorporate related information to improve the recommendation task. We propose to unify these approaches: using information from interactions in other domains as well as external knowledge graphs to make predictions in a new domain that would not be possible with either information source alone. We develop a new model and demonstrate the additive benefit of these approaches on a dataset derived from millions of users' queries for content across three domains (videos, music, and books) in a live virtual assistant application. We demonstrate significant improvement on overall recommendations as well as on recommendations for new users of a domain.
\end{abstract}

\begin{IEEEkeywords}
Conversational AI Agents, User Preference Learning, Knowledge Graphs, Graph Neural Networks
\end{IEEEkeywords}

\section{Introduction}
\label{sec:intro}

Recommender systems are prevalent from e-commerce to social media to video streaming. However, there are still challenging aspects to developing accurate recommender services. One of the main challenges is how to integrate new users. Since these systems rely on previous interaction data, they perform poorly when a user has little or no interaction history. There have been many approaches to alleviate this {\em cold start} issue from meta-learning \cite{maml}, to using side-information, to using data from other domains \cite{Zhu2021CrossDomainRC}. 

 Using data from multiple domains becomes especially important when the service needs to make recommendations in those other domains as well. This is known as multi-domain recommendation \cite{Zhu2021CrossDomainRC}. In multi-domain recommendation, the user representations are aligned through overlapping users, i.e. users who have previous interactions in both domains. Meanwhile, the item representations are only aligned indirectly through these users. For example, the item, \texttt{Book:Harry\_Potter} will only share some semblance to \texttt{Movie:Harry\_Potter} because the same users enjoy both the movie and the book. A conversational AI assistant offers an excellent use case for multi-domain recommendations, as it involves millions of real user interactions across various domain-specific content. Learning user preferences from historical multi-domain interactions is crucial for building a successful personalized AI assistant.

In a parallel line of work, researchers have been working on using knowledge graphs (KGs) to enhance recommender systems in the single domain setting \cite{Wang2019MultiTaskFL}. This recent line of work has found success using a multi-task approach, in which the knowledge graph is trained on a knowledge graph embedding objective and then linked to a recommender trained on a recommendation objective. 

Our research aims to use recent advances in KG enhancement to improve the efficacy of multi-domain recommendation in both the general setting and the zero-shot setting for new users. Since KG enhancement aligns item representations, it should have an additive benefit to multi-domain recommendation models that align representations over users.  

The contributions of this paper include 
\begin{enumerate}
    \item A novel state-of-the-art model combining multi-domain techniques and multi-task KG enhancement.
    \item Evaluation of the effect of KG enhancement with different base architectures on a real-world industry dataset, applied to a conversational AI assistant use case involving millions of users across three interaction domains. 
    \item Results that demonstrate KG enhancement is complementary to multi-domain models and offers significant improvement over baselines in both the general and zero-shot setting. 
    
\end{enumerate}

\section{Background}
\label{sec:background}
\textbf{User-item interaction through AI assistant} Users interact with the conversational agent by providing an input termed a ``\textit{query}''. Within the agent, there is a natural language understanding (NLU) component designed to comprehend the details of the given query, such as classifying the query into a domain, extracting and resolving entities from the query. This process is how we capture multi-domain user-item interactions with the conversational AI. Let $\gamma$ be an integer such that $1 \leq \gamma < \infty$. The natural language understanding of a query for our purpose can be understood as a mapping function, $h: Q \rightarrow D \times [E]^\gamma$, where $Q$ refers to the query space, $D$ refers to the domain space and $E$ refers to the entity space. The entity space, $E:= E_T \times E_V$, may further be decomposed into the entity type space $E_T$ and the entity value space $E_V$. As an example, given a query string $q=$``\textit{Play The Real Slim Shady.}'', the corresponding NLU hypothesis is $h(q)=$\textit{(Music, [(SongName, The Real Slim Shady)])} where the domain is \textit{Music} and the entity value is \textit{"The Real Slim Shady"} with the entity type of \textit{"SongName"}.

\textbf{Multi-Domain Recommendation}\ \ We define the multi-domain interaction graph $\mathcal{G} = (\mathcal{U}, \mathcal{V}, \mathcal{I}, \mathcal{D})$ where $\mathcal{D}$ is the set of domains, $\mathcal{U}$ is the set of users, $\mathcal{V} = \mathcal{V}_1\cup...\cup\mathcal{V}_D$ is the set of all items separable into $D$ domains, and $\mathcal{I}$ is the set of all observed user-item interactions in the graph. $\mathcal{I}_i=\{(u,v)|v\in \mathcal{V}_i, (u,v)\in\mathcal{I}\}$ is the set of interactions involving items from the $i$-th domain. The goal of multi-domain recommendation is given a user $u$ and a domain $d$, rank all the items in $\mathcal{V}_d$ such that items the user wants to interact with are ranked as highly as possible. This task is well aligned with user behavior as a user knows what domain they want to interact with (e.g. listen to music, read a book, etc.).  The zero-shot task is the exact same except that user $u$ has no prior interactions in $\mathcal{I}_d$.   

\textbf{Knowledge Graphs}\ \ Knowledge graphs (KGs) encode factual information about the world in a structured way. The most common formulation defines a finite set of entities and relation types and encodes the true relations between entities as triples in the form $(h,r,t)=(\textit{head}, \textit{relation}, \textit{tail})$. E.g. if the triple (\textit{Love Story}, \textit{performer}, \textit{Taylor Swift}) is in the database, then we know \textit{Taylor Swift} is the \textit{performer} of \textit{Love Story}. We consider a knowledge graph $\mathcal{K}=(\mathcal{E}, \mathcal{R}, \mathcal{T})$ where $\mathcal{E}$ is the set of entities, $\mathcal{R}$ is the set of relation types, and $\mathcal{T}$ is the set of triples of the form $(h,r,t)$ with $h,t \in \mathcal{E}$ and $r \in \mathcal{R}$.

\section{Related Works}
\label{sec:related_works}

\textbf{Multi-Domain Works}\ \ We study multi-target cross domain recommendation \cite{Zhu2021CrossDomainRC}, in which each domain is a target of recommendation and there can be more than two domains. \cite{Zhu2020AGA} introduces a dual attention model for dual domain recommendation and then expands on it to generalize to a greater number of domains \cite{ Zhu2021AUF}. \cite{Cui2020HeroGRAPHAH} uses graph neural networks as well as domain specific modules for the multi-domain setting. \cite{Guo2022DisentangledRL} expands on this by enforcing a notion of inter-domain agreement. Other approaches have used meta-learning to regularize the learning process to achieve multi-domain results \cite{Luo2022MAMDRAM}.

\textbf{Knowledge Graph Enhancement Works}\ \ Previous works use item features from knowledge graphs \cite{Zhao2018KB4RecAD}, but more recent works have sought to better integrate the KG via KG enhancement. This was first introduced in the multi-task learning format \cite{Wang2019MultiTaskFL}. Besides the recommendation objective, the other task is a knowledge graph embedding task such as TransE \cite{transE}. Since then a number of derivative works have looked at further advancements such as adding graph neural networks for aggregating user preferences \cite{Ma2021MNIAE, Zhang2021AKG} or for knowledge graph aggregation \cite{Wang2021EnhancedKG, Wang2021MultitaskFL} or both \cite{Xia_Huang_Xu_Dai_Zhang_Yang_Pei_Bo_2021} as well as other architectural perturbations \cite{Fan2022ImprovingRS}. All of these works explore the single domain setting.

\section{Model Architecture}
\label{sec:architecture}

\begin{figure*}
    \centering
    \includegraphics[trim={0 1cm 0 2cm},width=0.9\textwidth]{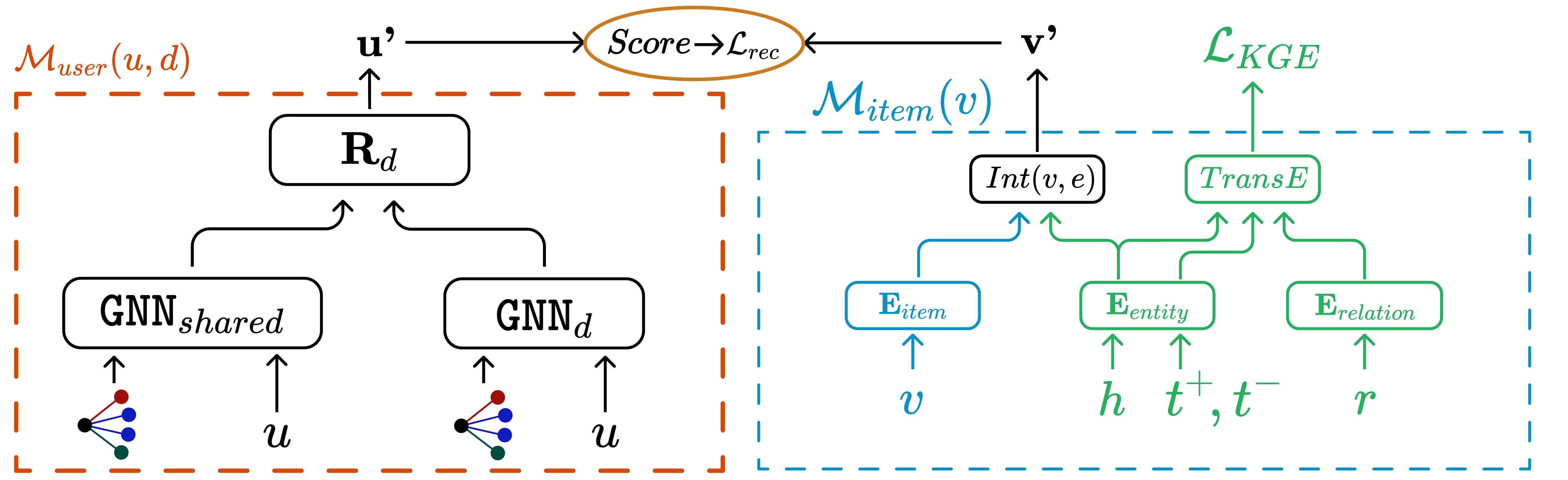}
    \caption{The combined MultiDomain KGE+GNN model provides recommendations for user $u$ based on their interaction graph and recommends items $v$ which are enhanced with KG embeddings using KG edges of the form $(h,r,t)$. KG \textcolor{ForestGreen}{entity} and \textcolor{ForestGreen}{relation} embeddings are trained using KG embedding task, while the \textcolor{Cerulean}{item} embeddings and interaction module are trained over the recommendation task. The user model contains one shared Graph Neural Network and multiple unique ones for each domain. Note that the domain specific GNNs still uses all prior interactions from multiple domains, but the weights and output is specific one domain.}
    \label{fig:KGE}
\end{figure*}

The model aims to produce a recommendation score of each item $v\in \mathcal{V}$ for user $u\in \mathcal{U}$. The model has three components: a user encoder $\mathcal{M}_{user}$, an item encoder $\mathcal{M}_{item}$, and a scoring function $f$. $\mathcal{M}_{user}$ generates an $h$ dimensional user embedding $\textbf{u}\in\mathbb{R}^h$. $\mathcal{M}_{item}$ generates item embedding $\textbf{v}\in\mathbb{R}^h$. The scoring function generates a scalar score for the relevance of an item to a user based on the embeddings, higher scores indicate a better item to recommend. Figure \ref{fig:KGE} shows the complete proposed model. 

In this work we always use inner-product as the scoring function: $f(\textbf{u},\textbf{v}) = \langle\textbf{u},\textbf{v}\rangle$

\subsection{Simple Embedding Baseline}


For the simple baseline we consider using a one hot embedding tables for items and users. Specifically, 
\begin{equation}
    \mathcal{M}_{user}(u) = \textbf{E}_{user}(u);\ \ \mathcal{M}_{item}(v) = \textbf{E}_{item}(v)
\end{equation}
where $\textbf{E}_{user}\in\mathbb{R}^{|\mathcal{U}|\times h}$ and $\textbf{E}_{item}\in\mathbb{R}^{|\mathcal{V}|\times h}$ are user and item embedding tables: Each row embeds a single user/item. 

The model is trained over batches of edges from the data. For each observed interaction $(u,v^+)$, we sample sets of negative pairs $(u,v^-)$ and train using a Margin Ranking Loss:

\begin{equation}
    \mathcal{L}_{rec} = max\left(0, f(\textbf{u}, \textbf{v}^-) - f(\textbf{u}, \textbf{v}^+) + \gamma_1\right) 
\end{equation}
where $\textbf{v}^+$ and $\textbf{v}^-$ are the item embeddings from $\mathcal{M}_{item}$ for the positive and negative samples, respectively. $\gamma_1$ is the margin hyperparameter. This loss attempts to ensure that positive scores are at least $\gamma_1$ greater than the negative scores. 

\subsection{Knowledge Graph Enhancement}

Item representations can often be improved by using an external KG. Recent works have done KG enhancement with multi-task training in which the model is trained on an auxiliary KG embedding task.

\textbf{KG Embedding}\ \ This work uses the TransE knowledge graph embedding task as its multi-task objective \cite{transE}. TransE defines the following distance metric $d$ for triples $(h,r,t)$:

\begin{equation}
    d(h,r,t) = \|\textbf{E}_{ent}(h) + \textbf{E}_{rel}(r) - \textbf{E}_{ent}(t)\|_2
\end{equation}

where $\textbf{E}_{ent}\in\mathbb{R}^{|\mathcal{E}|\times h}$ and $\textbf{E}_{rel}\in\mathbb{R}^{|\mathbb{R}|\times h}$ are the entity and relation embedding tables. The loss function is Margin Ranking Loss which ensures the distance metric for true triples is at least $\gamma_2$ less than for the negatively sampled ones

\begin{equation}
    \mathcal{L}_{KGE} = max\left(0, d(h,r,t^+) - d(h,r,t^-) + \gamma_2)\right)
\end{equation}

\textbf{Enhancement}\ \ To integrate the TransE embeddings for entity $e$ with the recommendation system's item $v$, we define an interaction block $Int(v, e)$ that uses the item embedding $\textbf{v}$ and entity embedding $\textbf{e}$ and outputs an updated item representation, $\textbf{v}'\in\mathbb{R}^h$. We use a multi-layer perceptron (MLP) for this component.
\begin{equation}
    \textbf{v}' = \textrm{Int}\left(\textbf{v},\textbf{e}\right)=\textrm{MLP}(\textbf{v} || \textbf{e}) 
\end{equation}
$\textbf{v}'$ is used in place of \textbf{v} for $L_{rec}$.

\begin{table}[!t]
\centering
\caption{Examples of dialogue sessions and extracted user-item interactions}
\label{tab:example}
\resizebox{.9\columnwidth}{!}{%
\begin{tabular}{@{}cl@{}}
\toprule
Domain & \multicolumn{1}{c}{Dialogue} \\ \midrule
\textbf{Music} &
  \begin{tabular}[c]{@{}l@{}}{[}USER{]}: play \textbf{hello} by \textbf{adele}.\\ {[}AGENT{]}: Okay hello by adele from Margie's Spotify.\end{tabular} \\ \midrule
\textbf{Video} &
  \begin{tabular}[c]{@{}l@{}}{[}USER{]}:  play \textbf{harry potter and the goblet of fire}.\\ {[}AGENT{]}: Okay. Playing Harry Potter and the Goblet of Fire.\end{tabular} \\ \midrule
\textbf{Books} &
  \begin{tabular}[c]{@{}l@{}}{[}USER{]}: Read the book \textbf{Harry Potter}.\\ {[}AGENT{]}: Resuming Greg's book: Harry Potter and the Goblet of Fire.\end{tabular} \\ \bottomrule
\end{tabular}%
}
\end{table}

\begin{table}[!t]
\caption{Statistics of training graph}
\centering
\begin{tabular}{lccc}
\hline
\textbf{} & \textbf{Music} & \textbf{Video} & \textbf{Books} \\
\hline
\#customer & 2.8M & 2.8M & 67K \\
\#items & 120K & 150K & 7K \\
\#edges & 25M & 11M & 91K \\
Sparsity & 99.9926\% & 99.9974\% & 99.981\% \\
\hline
\end{tabular}
\label{tab:stats}
\end{table}

\subsection{Multi Domain Recommendation}
To add multi-domain recommendation capabilities, we follow the Deep Sharing User Representations paradigm from \cite{Zhu2021CrossDomainRC} as exemplified by \cite{Lian2017CCCFNetAC}. We use a multi-domain variant of the simple embedding baseline. Here we define $D$ separate domain-specific user embedding tables $\textbf{E}_{user,d}$ and one shared embedding table $\textbf{E}_{user,shared}$. The final user embedding for a user interacting with domain $d$ is the concatenation of the shared and the domain-specific embeddings. 

\begin{equation}
    \mathcal{M}_{user}(u,d) = \textbf{R}_d\left(\textbf{E}_{user,shared}(u) || \textbf{E}_{user,d}(u)\right)
\end{equation}


\begin{table*}[h]
\caption{Overall results on recommendation tasks as measured by mean reciprocal rank (MRR) and hits at 100.}
\centering
\begin{tabular}{lcccccccc}
\hline
\textbf{Model} &  \multicolumn{2}{c}{\textbf{Overall}} & \multicolumn{2}{c}{\textbf{Music}} & \multicolumn{2}{c}{\textbf{Video}} & \multicolumn{2}{c}{\textbf{Books}}\\
 & \textbf{MRR} & \textbf{H@100} & \textbf{MRR} & \textbf{H@100}  & \textbf{MRR} & \textbf{H@100} & \textbf{MRR} & \textbf{H@100} \\
\hline
\textbf{SMF} \cite{rendle2012bpr} & 0.011 & 0.109 & 0.0171 & 0.1699 & 2E-5 & 0.0005 & 4E-5 & 7E-7 \\
\textbf{CAT-ART} \cite{Li2022OneFA} & 0.016  & 0.162 & 0.024 & 0.2446 & 0.0005 & 0.0144 & 2E-6 & 0.0001  \\
\hline
\textbf{Base Model} & 0.003 & 0.005 & 0.001 & 0 & 0.001 & 0 & 0.183 & 0.519 \\
\textbf{KGE} & 0.026 & 0.218 & 0.029 & 0.257 & 0.014 & 0.12 & \textbf{0.197} & 0.612 \\
\textbf{Multi-domain (MultD)} & 0.013 & 0.138 & 0.018 & 0.205 & 0.001 & 0.004 & 0.150 & 0.569 \\
\textbf{MultD KGE} & 0.027 & 0.239 & 0.028 & 0.26 & 0.013 & 0.117 & 0.158 & \textbf{0.628} \\
\textbf{MultD GNN} & 0.036 & 0.277 & 0.042 & 0.334 & 0.02 & 0.147 & 0.193 & 0.552 \\
\textbf{MultD KGE+GNN} & \textbf{0.037} & \textbf{0.279} & \textbf{0.044} & \textbf{0.342} & \textbf{0.021} & \textbf{0.153} & 0.195 & 0.472 \\
\hline

\end{tabular}
\label{tab:general_results}
\end{table*}

\begin{table*}[h]
\caption{Zero-shot results. Results on users who did not interact with the domain in the training set.}
\centering
\begin{tabular}{lcccccc}
\hline
\textbf{Model} & \multicolumn{2}{c}{\textbf{Music}} & \multicolumn{2}{c}{\textbf{Video}} & \multicolumn{2}{c}{\textbf{Books}}\\
 & \textbf{MRR} & \textbf{H@100}  & \textbf{MRR} & \textbf{H@100} & \textbf{MRR} & \textbf{H@100} \\
\hline
\textbf{Base Model} & 0.001 & 0 & 0.001 & 0 & 0.026 & 0.083 \\
\textbf{KGE} & 0.04 & 0.278 & 0.025 & 0.161 & 0.093 & \textbf{0.575} \\
\textbf{Multi-domain (MultD)} & 0.017 & 0.242 & 0.001 & 0.003 & 0.082 & 0.515 \\
\textbf{MultD KGE} & 0.026 & 0.313 & 0.023 & 0.148 & 0.12 & 0.563 \\
\textbf{MultD GNN} & 0.042 & 0.271 & 0.021 & 0.172 & 0.167 & 0.432 \\
\textbf{MultD KGE+GNN} & \textbf{0.043} & \textbf{0.317} & \textbf{0.025} & \textbf{0.176} & \textbf{0.179} & 0.494 \\
\hline

\end{tabular}
\label{tab:zeroshot_results}
\end{table*}

If a user has never interacted in domain $d$ then $\textbf{E}_{user,d}(u)$ will just be the random initialization but $\textbf{E}_{user,shared}(u)$ will still be meaningful. To leverage the combined information and shared correlations between the target domain and the other domains, and to reduce the dimension back to $h$, we use a domain-specific, two layer MLP, $\textbf{R}_d$, the weights of which are unique to the target domain, to interact the domain specific and cross-domain embedding (Similar to $Int(v,e)$). 

\textbf{Graph Neural Network}\ \ Graph neural networks (GNNs) are common in recommendation systems and were recently introduced for multi-domain settings \cite{Cui2020HeroGRAPHAH}. We replace user embedding tables with inductive GNNs. 

We replace $\textbf{E}_{user,shared}(u)$ with $\texttt{GNN}_{shared}(u, \mathcal{N}(u))$ where $u$'s neighborhood $\mathcal{N}(u)$ is the items user $u$ has interacted with. We likewise replace the domain-specific embedding tables $\textbf{E}_{user,d}(u)$ with domain-specific GNNs: $\texttt{GNN}_{d}(u, \mathcal{N}(u))$. The GNN aims to encode a user $u$ based on its neighbors. We use the GNN from \cite{Cui2020HeroGRAPHAH}, which uses a fast additive attention for node aggregation. This GNN iteratively learns an attention distribution over the neighbors of $u$. The initial item representations come from the item embedding table $\textbf{E}_{item}$. The initial user embedding comes from a domain embedding table $\textbf{E}_{d}\in\mathbb{R}^{|\mathcal{D}\times h|}$ rather than a user embedding table to make the GNN inductive over users. This forces the model to rely on the GNN to learn personalized user representations. Final user representation is:

\begin{equation}
    \mathcal{M}_{user}(u,d) = \textbf{R}_d\left(\texttt{GNN}_{shared}(u, \mathcal{N}(u)) || \texttt{GNN}_{d}(u, \mathcal{N}(u))\right)
\end{equation}

\section{Experiments}
\label{sec:experiments}

\subsection{Dataset}
We sample dialogue turns between users and a large scale conversational AI agent from de-identified historic interactions and keep three predefined business domains: Music, Video, Books (see Table \ref{tab:example}). We leverage the NLU component, as described in Section \ref{sec:background}, to extract user-item interaction history. We rely on machine-based dialogue evaluation metrics \cite{gupta2021robertaiq}, along with other rich signals such as music playback rate, weekly play frequency, and book purchase records, to ensure the user has a preference for the given item. We use an entity linking model \cite{Ayoola2022Refined} to link the item to Wikidata knowledge base. We further explore a manual verification process and apply additional rule-based filters to ensure that the accuracy of entity linking exceeds 96\%. We then construct an user/item interaction graph and keep the users who had interacted with at least two domains. For training, we use one month of interactions (over 30 million). For validation and testing, we use the following two weeks and discard repeat interactions that were seen in the training set. Statistics of the training graph are shown in Table \ref{tab:stats}. We use the filtered setting for our evaluation metrics in which known edges are removed from the comparisons \cite{transE}.

\subsection{Experimental Setup and Results}

In tables \ref{tab:general_results} and \ref{tab:zeroshot_results}, we show our approach against the base models. We compare six classes of models using the components described in section \ref{sec:architecture}. Each model is based off the simple baseline and can have the following variations: with and without knowledge graph enhancement (\textbf{KGE}), single-domain or \textbf{multi-domain}, and with and without the graph neural network (\textbf{GNN}). In order to facilitate fair experimental comparison, these models use the same objective and are optimized using grid search with similar search spaces for shared hyperparameters. We also compare against two off-the-shelf models. \textbf{SMF} is a single domain factorization method based on Bayesian Personalized Ranking \cite{rendle2012bpr}. \textbf{CAT-ART} is a recent state-of-the-art cross domain recommendation system using a contrastive autoencoder and an attention based transfer mechanism \cite{Li2022OneFA}.


\textbf{Effect of KG Enhancement.} We find that KG enhancement improves each model's performance across the board. Specifically, KGE versions outperform their corresponding non-KGE version in every case in both multi-domain and zero-shot settings. The impact is particularly pronounced for simpler models - adding KGE to the multi-domain model increases MRR from 0.013 to 0.027, representing a 108\% improvement. While the relative magnitude of improvement decreases as the base models become more sophisticated, KGE continues to provide meaningful gains even with the GNN model, particularly in the zero-shot setting where it improves Music H@100 from 0.271 to 0.317. This persistent benefit in the zero-shot scenario suggests that KG-enhanced item representations effectively capture cross-domain relationships, improving transfer learning between domains.


\textbf{Effect of Multi-Domain models.} We can look at the effect of adding multi-domain modeling capabilities by comparing the base single-domain vs multi-domain models as well as the single vs multi-domain models under KG enhancement. In both cases multi-domain confers a benefit, though the benefit is not as clearcut as adding KGE. Despite having the fewest interactions per user, the books domain sees least improvement from multi-domain learning as the task is the easiest. 

\textbf{Effect of GNN.} We observe that leveraging a GNN is a critical tool to improve recommendations. Specifically, our results indicate that the improvement in performance is unmistakable for all the baselines tested in almost all domains and metrics. 

\textbf{Zero-shot setting.} The trends observed in the general setting hold up in the zero-shot setting as well, though there is greater variation. The MultD KGE+GNN model is clearly the strongest model for users interacting with a new domain.

\textbf{Comparison to CAT-ART} We find significant improvement over CAT-ART. It is worth noting that CAT-ART was originally trained on interaction graphs with 1-3 orders of magnitude greater density and may not perform as well with sparser data.

\section{Conclusion}
\label{sec:conclusion}

We find that knowledge graph enhancement offers additive benefit to multi-domain recommendation, and the proposed multi-domain KGE + GNN model offers very strong recommendation performance on both returning and new users across multiple domains. Our proposed methods have been proven effective in a scalable, real-world AI assistant use case, bringing significant benefits to personalized AI assistants.

\bibliographystyle{IEEEtran}
\bibliography{refs}

\end{document}